\documentstyle[prl,aps,multicol,epsf]{revtex}
\input psfig
\begin {document}
\draft
\title {Theory of Weiss oscillations in the presence of small-angle
impurity scattering.} 
\author { A. D. Mirlin$^*$ and P. W\"olfle}
\address{Institut f\"ur Theorie der kondensierten Materie,
Universit\"at Karlsruhe, 76128 Karlsruhe, Germany}
\date {February 26, 1998}
\maketitle
\begin {abstract}
We calculate the magnetoresistivity of a two-dimensional
electron gas in the presence of a periodic potential within classical
transport theory, using realistic models of impurity scattering. The
magnetooscillations induced by geometric resonance of the cyclotron
orbits in the periodic grating, known as Weiss oscillations,
are shown to be affected strongly by
the small-angle scattering processes dominant in conventional
semiconductor heterostructures. Our results are in full agreement with
experimental findings.

\end {abstract}
\begin{multicols}{2}
\section{Introduction}
\label{s1}

In 1989, Weiss, von Klitzing, Ploog, and Weimann \cite{weiss}
discovered the striking effect of a periodic potential varying in one
direction on the magnetoresistivity of a two-dimensional electron gas
(2DEG). They observed that even a weak potential modulation (grating)
with a wave vector ${\bf q}\parallel {\bf e}_x$ induces very strong
oscillations of the magnetoresistivity $\rho_{xx}(B)$, while showing
almost no effect on $\rho_{yy}(B)$ and $\rho_{xy}(B)$. 
The oscillation period corresponds
to a change of the ratio $2R_c/a$ by unity, where $R_c$ is the
cyclotron radius, and $a=2\pi/q$ the grating wave length. These
findings were explained by Gerhardts {\it et al} \cite{gerhardts89}
and  Winkler {\it et al} \cite{winkler} in terms of the
quantum-mechanical band structure induced by the modulation
(see also \cite{peeters} for explanation of the weak
oscillations induced by the grating in $\rho_{yy}(B)$ and
$\rho_{xy}(B)$ and \cite{peeters,beton} for discussion of the
temperature dependence of the magnetoresistivity).
Beenakker \cite{beenakker} presented a semiclassical theory of the
effect. He  showed that the geometric resonance of the cyclotron
motion in the grating induces an extra contribution to the
drift velocity of the guiding center, with a r.m.s. amplitude 
 oscillating with period $2R_c/a$. These arguments were supported
by an exact solution of the Boltzmann equation, where isotropic
scattering with a relaxation time $\tau$ was assumed. However, while
accounting for the above experimental features, the
theoretical results \cite{gerhardts89,winkler,beenakker} 
disagree strongly with the experiment as far as
the damping of the oscillations with decreasing  magnetic
field is concerned. 
Specifically, in the regime $\omega_c\tau\gg 1$ (which is
valid for the experimental value of the transport time $\tau$ in the
whole range of the magnetic fields where the oscillations are
observed) the theory \cite{gerhardts89,winkler,beenakker} 
predicts a linear decrease of
the oscillation amplitude, $\Delta\rho_{xx}(B)\propto B$:
\begin{equation}
{\Delta \rho_{xx}\over\rho_{xx}}\simeq \eta^2{l^2\over a
R_c}\cos^2(qR_c-\pi/4)
\label{e1}
\end{equation}
Here $\eta$ is the relative amplitude of the potential modulation 
$eU(x)=-\eta E_F\cos qx$, $l=v_F\tau$ is the  mean free path, 
$\omega_c=eB/m=v_F/R_c$ the cyclotron frequency, $E_F$ the Fermi
energy, and $v_F$ the Fermi velocity. In contrast with this result, the
experimental damping is much stronger, so that only few oscillations
($6-7$ in \cite{weiss}) are observed. 

The problem of the damping of the Weiss oscillations was addressed in
a number of publications. In Ref.\cite{beton} an empirical
observation was made
that inclusion of the factor $\exp(-\pi/\omega_c\tau_s)$ in
Eq.(\ref{e1}) with $\tau_s^{-1}$ being the total relaxation rate (as
opposed to the momentum relaxation rate $\tau^{-1}$) yields a
reasonable description of the experimentally observed damping. The
authors of Ref.\cite{paltiel} argued quite oppositely that only the
transport 
time should enter the expression for the damping factor, and proposed
an empirical form $\exp[-4\pi^3\gamma^2(R_c/a)^2(\omega_c\tau)^{-1}]$,
with a 
fitting parameter $\gamma$. Monte Carlo simulations \cite{boggild}, as
well as numerical solution of the Boltzmann
equation \cite{gerhardts}, showed that allowing for 
anisotropic scattering leads to a stronger damping of the
oscillations. However, despite considerable interest in this problem,
no quantitative theory of the Weiss oscillations for the
experimentally relevant situation, when  small angle scattering is
dominant, has been available so far. The purpose of the present paper
is to fill in this gap.

\section{Boltzmann equation formalism}
\label{s2}

We begin by reminding the reader briefly about
the Boltzmann equation
approach to the problem \cite{beenakker}. One starts from the kinetic
equation for the distribution function $F(x,{\bf n})$ of electrons
on the Fermi surface,
\begin{eqnarray}
&&{\cal L} F(x,{\bf n}) = -e v(x) {\bf En}\ , \nonumber\\
&&{\cal L}=v(x){\bf n}{\partial\over\partial {\bf r}} +
\omega_c{\partial\over\partial\phi}-\sin\phi
v'(x){\partial\over\partial\phi} - C\ , \label{e2}
\end{eqnarray}
where ${\bf n}=(\cos\phi,\sin\phi)$ is the direction of the electron
velocity, $v(x)=[2m(E_F+eU(x))]^{1/2}$ is the magnitude of the
velocity, $-e$ is the electron charge, and $C$ is the collision
integral (to be discussed below).
In the absence of the grating, the solution is given by
\begin{equation}
F^{(0)}({\bf n})=-{2\over e\nu v_F} n_i\sigma^{(0)}_{ij}E_j\ ,
\label{e3}
\end{equation}
where $\hat{\sigma}^{(0)}$ is the Drude conductivity tensor,
\begin{equation}
\hat{\sigma}^{(0)}={\sigma_0\over 1+S^2}\left(
\begin{array}{cc}
1        &      -S \\
S        &       1 
\end{array}
\right)\ ;\qquad \sigma_0=e^2\nu v_F^2\tau/2\ ,
\label{e4}
\end{equation}
$S=\omega_c\tau$, and
$\nu$ is the density of states at the Fermi level. Using the
Ansatz $F=[v(x)/v_F]F^{(0)}+F^{(1)}$, one may transform
 Eq.(\ref{e4}) into the form 
\begin{equation}
{\cal L}F^{(1)}(x,{\bf n})=v(x)v'(x){e\tau\over 1+S^2}(E_x-SE_y)
\label{e5}
\end{equation}
This implies that the grating induced correction to the conductivity
tensor, $\hat{\sigma}=\hat{\sigma}^{(0)}+\delta\hat{\sigma}$, is given
by
\begin{equation}
\delta\hat{\sigma}={\sigma_0\over 1+S^2}\langle
G(x,\phi)v(x)\cos\phi\rangle \left(
\begin{array}{cc}
1        &      -S \\
S        &       -S^2 
\end{array}
\right)\ ,
\label{e6}
\end{equation}
where $G(x,\phi)$ is the solution to the equation
\begin{equation}
{\cal L}G=-2v(x)v'(x)/v_F^2\equiv - 2 eU'(x)/mv_F^2
\label{e7}
\end{equation}
and the angular brackets $\langle\ldots\rangle$ denote the averaging
over both the velocity direction and a period of the modulation,
\begin{equation}
\langle{\cal O}(x,\phi)\rangle={1\over a}\int_0^a dx
\int_0^{2\pi}{d\phi\over 2\pi}{\cal O}(x,\phi).
\label{e8}
\end{equation}
Calculating now the resistivity tensor $\hat{\rho}=\hat{\sigma}^{-1}$,
one finds that $\hat{\rho}=\hat{\rho}^{(0)}+\delta\hat{\rho}$, where
$\hat{\rho}^{(0)}=[\hat{\sigma}^{(0)}]^{-1}$ and the only non-zero
component of $\delta\hat{\rho}$ is $\delta\rho_{xx}$ given by
\begin{equation}
\delta\rho_{xx}=-{1+S^2\over\sigma_0} {\langle Gv\cos\phi\rangle \over 1+ 
\langle Gv\cos\phi\rangle}\ .
\label{e9}
\end{equation}
For a weak modulation, $\eta\ll 1$, the corrections  $\delta\hat{\rho}$
and $\delta\hat{\sigma}$ are proportional to $\eta^2$. Therefore, to
calculate the average $\langle Gv\cos\phi\rangle$ 
in the leading order, one would have to solve the
Boltzmann equation (\ref{e7}) up to the second order in $\eta$. 
However, it was observed by Beenakker \cite{beenakker}  that the
following identity holds:
\begin{equation}
\langle G(x,\phi)v(x)\cos\phi\rangle={e\tau\over m(1+S^2)}
\langle G(x,\phi)U'(x)\rangle\ .
\label{e10}
\end{equation}
Since $U(x)\propto\eta$, it is now sufficient to calculate $G(x,\phi)$
to first order in $\eta$, in order to find $\delta\hat{\sigma}$ or
$\delta\hat{\rho}$ to the  order $\eta^2$ . We note that, while only
presented for the isotropic scattering case  in Ref.\cite{beenakker},
the identity (\ref{e10}) holds for a general  collision
integral $C$ as well, since its derivation uses only the fact that
$C\{1\}=0$, $C\{\cos\phi\}=-(1/\tau)\cos\phi$, 
 and $C\{\sin\phi\}=-(1/\tau)\sin\phi$. 

To proceed further, one has to solve the kinetic equation (\ref{e7}).
This requires first specifying the collision integral. Its general form
reads:
\begin{equation}
C\{F\}(\phi)=\int d\phi' P(\phi-\phi')[F(\phi')-F(\phi)]\ ,
\label{e11}
\end{equation}
where $P(\phi-\phi')$ is the scattering cross-section. The collision
integral can be most conveniently characterized by its eigenvalues
\begin{eqnarray}
&&C\{e^{im\phi}\}=-\lambda_m e^{im\phi}\ ;\qquad m=0,\pm 1,\pm2,\ldots
\nonumber\\ 
&&\lambda_m=\lambda_{-m}=\int d\phi P(\phi) (1-\cos m\phi)\ge 0\ .
\label{e12}
\end{eqnarray}
Note that $\lambda_0=0$ reflects the particle number conservation;
$\lambda_1=1/\tau$ is the transport relaxation rate, while
$\lambda_\infty=1/\tau_s$ is the total relaxation rate. Expanding the 
solution of Eq.(\ref{e7}) in the angular harmonics, 
$G(x,\phi)=\sum_{m=-\infty}^\infty g_m e^{iqx} e^{im\phi} +c.c.$, we
get  to the leading order in $\eta$ the following
coupled system of equations \cite{MW}:
\begin{eqnarray}
&&a_ng_n=g_{n+1}+g_{n-1}+c_n\ ;  \label{e13}\\
&&a_n={2i\over v_F q}(in\omega_c+\lambda_n)\ ;\qquad c_n=-{\eta\over
v_F}\delta_{n0} \nonumber
\end{eqnarray}
According to Eqs.(\ref{e9}), (\ref{e10}), the correction to the
resistivity tensor is determined by $g_0$, since
\begin{equation}
\langle Gv\cos\phi\rangle=-{\eta q E_F \tau\over m
(1+S^2)}\mbox{Im}\,g_0.
\label{e13a}
\end{equation}
In turn, $g_0$ is found from the system (\ref{e13}) as 
\begin{equation}
g_0={\eta\over v_F}{1\over R_1+ R_{-1}}\ ,
\label{e14}
\end{equation}
where $R_1=g_1/g_0$ is determined by the set of  (homogeneous)
equations (\ref{e13}) with $n\ge 1$, and
$R_{-1}=R_1|_{\omega_c\to-\omega_c}$. For a generic set of eigenvalues
$\lambda_n$, the solution $R_1$ can be presented as a continuous
fraction,
\begin{equation}
R_1={1\over a_1-{1\over a_2-{1\over a_3-\ldots}}} \ .
\label{e15}
\end{equation}
The result can be expressed in closed form
 in the case when $\lambda_n$ depends linearly
on $n$, $\lambda_n=\tau^{-1}[1+2p(|n|-1)]$ with an arbitrary $p$,
yielding \cite{MW}
\begin{eqnarray}
&&R_1=-J_{1-}/J_-\ ;\qquad R_{-1}=J_{1+}/J_+\ , \label{e16}\\
&&J_{\pm}=J_{\pm(1-2p)i/(S\pm 2pi)}(Q/(1\pm 2pi/S))\ ; \nonumber\\
&&J_{1\pm}=J_{1\pm(1-2p)i/(S\pm 2pi)}(Q/(1\pm 2pi/S))\ , \nonumber
\end{eqnarray}
where $Q=qR_c$. 
The two most important particular cases are $p=0$, corresponding to
a white-noise random potential (isotropic scattering), and $p=1$,
describing the scattering by a white-noise random magnetic field with
the scattering rate $P(\phi)=\tau^{-1}\cot^2(\phi/2)$. In the next
section we investigate $\delta\rho_{xx}$ in these two situations,
exploiting the existence of the exact analytical solution
(\ref{e16}). This will give us a first example of how the small angle
scattering (dominant in the random magnetic field case, when
$\tau_s^{-1}=\infty$) modifies the magnetic field dependence of the
amplitude of the Weiss oscillations.

\section
{White-noise random potential versus random magnetic field}
\label{s3}
Using Eqs. (\ref{e13a}), (\ref{e14}), and (\ref{e16}), we find for
arbitrary values of the parameter $p$
\begin{equation}
\langle Gv\cos\phi\rangle=-{\eta^2 q l\over 2(1+S^2)} \mbox{Im}
\left({J_{1+}\over J_+}-{J_{1-}\over J_-}\right)^{-1}
\label{e18}
\end{equation}
For the particular case of isotropic scattering, $p=0$, the following
identity can be  used to simplify Eq.(\ref{e18}):
\begin{equation}
J_-J_{1+}-J_+J_{1-}={2i\over SQ}\left[J_+J_- -
{\sinh(\pi/S)\over\pi/S}\right]\ . 
\label{e19}
\end{equation}
This allows us to reduce Eq.(\ref{e18}) to the form
\begin{equation}
\langle Gv\cos\phi\rangle=-{\eta^2(ql)^2\over 4(1+S^2)}{J_+J_-\over 
{\sinh(\pi/S)\over\pi/S}-J_+J_-}\ .
\label{e20}
\end{equation}
Substituting (\ref{e20}) in (\ref{e9}) and assuming that $\eta$ is
small enough so that $\langle Gv\cos\phi\rangle\ll 1$ (this will be
valid for all $B$, if   $\eta^2 ql/4\ll 1$), we finally get
\begin{eqnarray}
\delta\rho_{xx}\over\rho_0 & = & {\eta^2\over 4}(ql)^2 {J_+J_-\over 
{\sinh(\pi/S)\over\pi/S}-J_+J_-} \nonumber\\
& \equiv & {\eta^2\over 4} (QS)^2 {J_{i/S}(Q)J_{-i/S}(Q)\over
{\sinh(\pi/S)\over\pi/S}-J_{i/S}(Q)J_{-i/S}(Q)}\ , \label{e21}
\end{eqnarray}
where $\rho_0=\sigma_0^{-1}$ is the Drude resistivity. Eq.(\ref{e21})
is completely equivalent to Beenakker's result (Eq. (5) of
Ref.\cite{beenakker}), which is, however, presented in \cite{beenakker} 
in the somewhat different form of an infinite series. The equivalence
follows from the Bessel function identity \cite{prudnikov}
\begin{equation}
\sum_{n=-\infty}^\infty {J_n^2(z)\over \nu^2+n^2}={\pi\over\nu\sinh\pi\nu}
J_{i\nu}(z)J_{-i\nu}(z)\ .
\label{e22}
\end{equation} 

We will assume that $ql\equiv QS\gg 1$, which is a necessary condition
for the existence of oscillations. At $Q\equiv qR_c\gg 1$ (i.e. in the
oscillation region), the asymptotic behavior of
$\delta\rho_{xx}/\rho_0$ reads according to Eq. (\ref{e21}) 
\begin{eqnarray}
{\delta\rho_{xx}\over\rho_0} & = & {\eta^2\over 2\pi} QS^2\cos^2(Q-\pi/4)\
,\qquad \pi/S\ll 1 \label{e23}\\
{\delta\rho_{xx}\over\rho_0} & = & {\eta^2\over 4} QS [1+2 e^{-\pi/S}
\sin 2Q]\ ,\qquad \pi/S\gg 1 \label{e24}
\end{eqnarray}
Thus, the amplitude of oscillations depends linearly on
 magnetic field $B$ in the high-field regime $\omega_c\gg
\pi/\tau$, and gets exponentially damped in $1/B$,
$(\delta\rho_{xx}/\rho_0)_{osc}\propto e^{-\pi/\omega_c\tau}$, in the
lowest fields, $\omega_c\ll\pi/\tau$. Note that for a typical value of
the transport time $\tau$ in the GaAs heterostructures, $\tau\sim
100\mbox{ps}$, the condition $\pi/\omega_c\tau\sim 1$ corresponds to
very low fields, $B\sim 0.01 \mbox{T}$.

We turn now to the random magnetic field scattering, 
where $p=1$. Analyzing
Eq. (\ref{e18}) in this case, we find the following two asymptotic
regimes:
\begin{eqnarray}
{\delta\rho_{xx}\over\rho_0} & = & {\eta^2\over 8} S^2\cos^2(Q-\pi/4)\
,\qquad 4Q/S\ll 1 \label{e25}\\
{\delta\rho_{xx}\over\rho_0} & = & {\eta^2\over 4} QS [1+2 e^{-4Q/S}
\sin 2Q]\ ,\qquad 4Q/S\gg 1 \label{e26}
\end{eqnarray}
Again Eq.(\ref{e25}) corresponds to a power-law behavior of the
oscillation amplitude in the range of relatively high magnetic fields $B$,
while Eq.(\ref{e26}) describes the exponential damping of the
oscillations in low fields. There is however a number of important
differences between the results for the isotropic scattering
[Eqs.(\ref{e23}), (\ref{e24})] and those for the random magnetic field
scattering [Eqs.(\ref{e25}), (\ref{e26})]:
\begin{itemize}
\item[i)] the ``high'' field behavior (\ref{e25}) is
$\delta\rho_{xx}/\rho_0\propto B^2$, rather than 
$\delta\rho_{xx}/\rho_0\propto B$ as in Eq.(\ref{e23});
\item[ii)] the value of the magnetic field, at which the exponential
damping starts, is much larger in the random magnetic field case
($\omega_c\tau\sim 2\sqrt{ql}\gg 1$), than for the isotropic
scattering  ($\omega_c\tau\sim\pi$);
\item[iii)] the low-field damping is of a Gaussian form 
(with respect to $1/B$) in the random magnetic field case,
$(\delta\rho_{xx}/\rho_0)_{osc}\propto\exp[-4ql/(\omega_c\tau)^2]$,
rather than of a simple exponential, $\exp(-\pi/\omega_c\tau)$, as in
Eq.(\ref{e24}). 
\end{itemize}
All these features are consequences of the small-angle scattering,
dominant in the random magnetic field case. In the next two sections
we will consider the Weiss oscillations in
the long-range correlated random potential, which also leads to 
small-angle scattering. Since in this case the exact solution can not
be expressed in a simple analytic form, we will use different methods 
to find the shape of the Weiss oscillations.

\section{Long-range random potential:
exponential damping of Weiss oscillations}
\label{s4}

The random potential experienced by electrons of a 2DEG in the GaAs
heterostructures is very smooth, since the impurities are separated
from the 2DEG layer by a relatively large distance (spacer)
$d_s>k_F^{-1}$ ($k_F$ being the Fermi wave vector). Assuming that the
positions of ionized donors are uncorrelated, while their
density is equal to that of the electrons, $n_e=k_F^2/2\pi$, and
treating the screening in the random phase approximation, one finds
for the random potential correlation function $W(|{\bf r}-{\bf
r'}|)=\langle U({\bf r})U({\bf r'})\rangle$ in momentum space
\begin{equation}
\tilde{W}(q)={n_e\over (2\nu)^2}e^{-2qd_s}\ ,
\label{e27}
\end{equation}
where $\nu=m/2\pi$ is the density of states per spin orientation. 
The scattering cross-section in the Born approximation is thus given
by
\begin{equation}
P(\phi)=\nu\tilde{W}(2k_F\sin\phi/2)={\pi n_e\over 2m}e^{-4k_F
d_s\sin\phi/2}\ . 
\label{e28}
\end{equation}
The total ($\tau_s^{-1}$) and the momentum ($\tau^{-1}$)
relaxation rates are defined as
\begin{eqnarray}
1/\tau_s &=& \int_0^{2\pi}d\phi P(\phi)={k_F\over 4md_s}\ ; \label{e29}\\
1/\tau &=& \int_0^{2\pi}d\phi P(\phi) (1-\cos\phi)={1\over 16 m k_F
d_s^3}\ ,
\label{e30}
\end{eqnarray}
so that their ratio is $\tau/\tau_s=(2k_F d_s)^2\gg 1$. Both $\tau$
and $\tau_s$ are experimentally measurable: the former is found from
the {\it dc} conductivity at $B=0$, while the latter follows
from the Dingle
analysis of the damping of the Shubnikov--de Haas oscillations. For
the structures of the type used in the Weiss oscillation experiment
\cite{weiss} ($n_e\sim 3\times 10^{11}\mbox{cm}^{-2}$, $d_s\sim 30
\mbox{nm}$), their typical values are $\tau\sim 50 \mbox{ps}$,
$\tau_s\sim 3\mbox{ps}$, so that the ratio $\tau/\tau_s\sim 20$ is
large, in agreement with the theory. Note, however, that the
experimental values of $\tau$ and $\tau_s$ are usually somewhat different
from the theoretical estimates (\ref{e29}), (\ref{e30}). This was
attributed to correlations between the positions of the 
scatterers \cite{coleridge,buks}.
Below we will express our results for the damping of the
Weiss oscillations 
in terms of $\tau$ and $\tau_s$, so that for comparison with experiment
one should simply use the experimental values of these quantities. 

As was shown in \cite{beenakker}, the physical origin of the
oscillations can be traced to 
the additional contribution to the drift velocity
of the guiding center,
\begin{equation}
\label{e31}
v_{dr}={1\over 2\pi B}\int_0^{2\pi}d\phi E(X+R_c\cos\phi)\ ,
\end{equation}
where $X$ is the $x$-coordinate of the guiding center and
$E(x)=-U'(x)=-(\eta E_F q/e)\sin qx$ is the electric field induced by
the grating.  Averaging $v_{dr}^2$ over $X$, one finds
\cite{beenakker}:
\begin{equation}
\label{e32}
\langle v_{dr}^2\rangle={1\over 2}\left( {\eta E_F q \over 2\pi e
B}\right)^2 \mbox{Re} \int_0^{2\pi} d\phi_1 d\phi_2 
e^{iqR_c(\sin\phi_1-\sin\phi_2)}\ .
\end{equation}
For $qR_c\gg 1$, the oscillations are determined by the interference
of the contributions from the vicinities of the saddle points
$\phi_1=0$, $\phi_2=\pi$ (and vice versa), leading to
\begin{equation}
\label{e33}
\langle v_{dr}^2\rangle_{osc} \propto \cos(2qR_c-\pi/2)
\end{equation}
(in this section we do not keep power-law prefactors, since we
calculate the exponential damping factor only).

Impurity scattering leads to a change $\Delta X$ of the guiding
center position $X$ during the time an electron moves along the
cyclotron orbit from $\phi=0$ to $\phi=\pi$. This introduces in
Eq.(\ref{e33}) a damping factor
\begin{equation}
\label{e34}
D=\langle e^{iq\Delta X}\rangle
\end{equation}
For isotropic scattering, any scattering event kicks an electron
far away from its original cyclotron orbit, so that $D$ is determined
simply by the probability for an electron to remain unscattered during
the half cyclotron period time $\pi/\omega_c$, yielding $D\simeq
e^{-\pi/\omega_c\tau}$, in agreement with Eq.(\ref{e24}). 
The situation is, however, different when the scattering is of
small-angle character, so that even after one or several scattering
events an electron remains close to its original cyclotron orbit and
contributes to the oscillations.

We first consider the random magnetic field scattering. Since in this
case the exact solution has been found (Eq.(\ref{e26}), this will be
an additional check for the correctness of our method of calculation
in this section.

In this case, the probability for an electron to be scattered by an
angle $\delta \phi$ in a small time interval $\delta t$  is given by
\cite{MAW} 
\begin{equation}
\label{e35}
{\cal P}(\delta\phi)={1\over \pi}{2\delta t/\tau\over
(2\delta t/\tau)^2+(\delta \phi)^2}
\end{equation}
Scattering by the angle $\delta\phi$ changes the position $X$ of the
guiding center by $R_c\sin\phi\delta\phi$, so that
\begin{equation}
\label{e36}
\Delta X=R_c\sum_k \sin\phi_k\delta\phi_k\ ,
\end{equation}
where $\phi_k\equiv\phi(t_k)=\omega_c t_k$; $t_k=k\pi/N\omega_c$;
$k=1,2,\ldots,N$; $N\to\infty$. According to Eq.(\ref{e35}),
\begin{equation}
\langle e^{iz\delta\phi}\rangle =\int d(\delta\phi){\cal
P}(\delta\phi)e^{iz\delta\phi} = e^{-2z\delta t/\tau}\ .
\label{e37}
\end{equation}
Therefore, the damping factor $D$ is equal to
\begin{eqnarray}
 D & =& \prod_k\langle e^{iqR_c\sin\phi_k\delta\phi_k}\rangle=\prod_k
e^{-2qR_c\sin\phi_k\delta t_k/\tau} \nonumber \\
  & = & \exp\left\{-{2qR_c\over
\omega_c\tau}\int_0^\pi d\phi\sin\phi\right\}
= \exp\left\{-{4qR_c\over \omega_c\tau}\right\}\ ,
\label{e38}
\end{eqnarray}
in precise agreement with the exact solution, Eq.(\ref{e26}). 

We turn now to the long-range random potential. In this case, the
scattering cross-section (\ref{e28}) implies that the distribution
function of the scattering angles $\delta\phi$ in a time interval
$\delta t$ is given by
\begin{equation}
\label{e39}
{\cal P}(\delta\phi)=\left(1-{\delta
t\over\tau_s}\right)\delta(\delta\phi)+ 
\delta t\, P(\delta\phi)\ .
\end{equation}
Therefore,
\begin{equation}
\langle e^{iz\delta\phi}\rangle = 1 - {\delta t \over\tau_s} {z^2\over
z^2+(2k_F d_s)^2}
\label{e40}
\end{equation}
This leads to the following result for the damping factor
\begin{eqnarray}
 D & =& \prod_k\langle e^{iqR_c\sin\phi_k\delta\phi_k}\rangle \nonumber\\
 &=&\exp\left\{-{(qR_c)^2\over \omega_c\tau_s}\int_0^\pi
{d\phi\sin^2\phi\over (qR_c\sin\phi)^2+(2k_Fd_s)^2}\right\}
\nonumber\\ 
& = &\exp\left\{-{\pi\over \omega_c\tau_s}\left[1-
{1\over\sqrt{1+(\tau_s/\tau)(qR_c)^2}}\right]\right\}\ .
\label{e41}
\end{eqnarray}
As is seen from Eq.(\ref{e41}), the region of  exponentially damped
oscillations can be further subdivided into two regimes:
$D\simeq\exp\{-\pi(qR_c)^2/2\omega_c\tau\}$ at $(qR_c)^2\ll
\tau/\tau_s$ and $D\simeq\exp\{-\pi/\omega_c\tau_s\}$ at $(qR_c)^2\gg
\tau/\tau_s$. In the former the leading contribution to $D$ comes from
multiply scattered carriers, while in the latter $D$ is governed by
those electrons, which remain unscattered within the time
$\pi/\omega_c$. 

The exponential damping found in the above
is efficient at low enough magnetic
fields, $\omega_c\tau<[(\pi/2)(ql)^2]^{1/3}$. At stronger magnetic
fields, the $B$-dependence of $\delta\rho_{xx}$ is of a power-law
nature. This will be considered in the next section
using a different method. 

\section{Long-range random potential: High
magnetic fields}
\label{s5}

To find the grating-induced correction to the resistivity at relatively
strong fields, we return to the Boltzmann equation approach of
Sec.\ref{s2}. For the scattering cross-section (\ref{e28}) the
eigenvalues $\lambda_l$ of the collision integral are equal to
\begin{eqnarray}
\lambda_l & = & \pi {n_e\over m}\int _0^\pi d\phi e^{-4k_F
d_s\sin\phi/2}(1-\cos l\phi) \nonumber\\
& \simeq &\pi {n_e\over m}\int _0^\pi d\phi e^{-2k_F
d_s\phi}(1-\cos l\phi) \nonumber\\
& = & \tau^{-1} {l^2\over (\tau_s/\tau)l^2 +1}
\label{e42}
\end{eqnarray}
The approximation we will make at this point will be to consider the
limit $\tau_s/\tau\to 0$, which reduces Eq.(\ref{e42}) to
$\lambda_l=l^2/\tau$. This approximation is valid for not too small
magnetic fields, such that $\omega_c\gg qv_F(\tau_s/\tau)^{1/2}$, 
where the total relaxation rate $\tau_s$ is
irrelevant, and only the transport one, $\tau$, is important. This
will allow us to find $\delta\rho_{xx}$ in the high-field regime (where
the $B$-dependence of the oscillation amplitude is of the power-law
nature), which was outside the regime of validity of the consideration
in Sec.\ref{s4}, where only the exponential damping factor was
studied. Furthermore, we will also reproduce the result of Sec.\ref{s4} 
for the intermediate regime
$(2/\pi)\omega_c\tau\ll(qR_c)^2\ll\tau/\tau_s$, where  the damping is
already exponential, but $\tau_s$ is still irrelevant.

Within the above approximation, the collision integral $C\{G\}$ is
replaced by the differential operator $\tau^{-1}\partial^2
G/\partial^2\phi^2$, and Eq.(\ref{e7}) with
$G(x,\phi)=g(\phi)e^{iqx}+c.c.$ takes the form
\begin{equation}
{\partial g \over\partial\phi}+iqR_c\cos\phi g-{1\over \omega_c\tau}
{\partial^2 g \over\partial\phi^2} = i\eta {q\over 2\omega_c}\ .
\label{e43}
\end{equation}
For $\omega_c\tau\gg 1$, $ql\gg 1$, we can solve this equation
iteratively considering the $\partial^2 g /\partial\phi^2$ term
as a perturbation. Dropping this term in (\ref{e43}) and
differentiating with respect to $\phi$ we find
\begin{equation}
{\partial^2 g \over\partial\phi^2}\simeq iqR_c \left(
g\sin\phi- {\partial g \over\partial\phi}\cos\phi\right)
\label{e44}
\end{equation}
Substituting this back to Eq.(\ref{e43}), we get a first order
differential equation, the solution of which reads
\begin{eqnarray}
g(\phi)& \simeq & {iq\eta\over 2\omega_c}\int_{-\infty}^\phi
d\phi' \nonumber \\
&\times & \exp\left[-iQ(\sin\phi-\sin\phi')-
{Q^2\over 2S}(\phi-\phi')\right]\ .
\label{e45}
\end{eqnarray}
Here we introduced again the compact notations $Q=qR_c$,
$S=\omega_c\tau$. 
Substituting now the solution (\ref{e45}) into Eqs. (\ref{e9}),
(\ref{e14}), we finally get
\begin{equation}
{\delta\rho_{xx}\over\rho_0}={\eta^2\over 2}S^2{\pi Q^2/2S\over \sinh 
(\pi Q^2/2S)}J_{iQ^2/2S}(Q)J_{-iQ^2/2S}(Q)\ .
\label{e46}
\end{equation}
Combining the results of this and the preceding section, we get a
complete description of $\delta\rho_{xx}$ in the full range of 
magnetic fields $B$, which will be analyzed in Sec.\ref{s6}. 

\section{Long-range random potential: Analysis of the results}
\label{s6}

According to the results of Secs. \ref{s4}, \ref{s5}, we find in general the
following three regions of the magnetic field $B$ with different
behavior of the oscillation amplitude:
\begin{itemize}
\item[i)] high field domain, $Q\ll Q_1$, with $Q_1=(2ql/\pi)^{1/3}$.
In this region $\pi Q^2/2S\ll 1$, and Eq.(\ref{e46}) yields
\begin{equation}
{\delta\rho_{xx}\over\rho_0}=\eta^2{S^2\over \pi Q}\cos^2(Q-\pi/4)
\label{e47}
\end{equation}
The amplitude of the  oscillations is proportional to $B^3$ in this
regime, in 
contrast to the linear  behavior, Eq.(\ref{e23}), for the
isotropic random potential, and to the $B^2$ behavior, Eq.(\ref{e25}) in
the case of random magnetic field scattering. 
\item[ii)] intermediate domain $Q_1\ll Q\ll Q_2$, with
$Q_2=\sqrt{2\tau/\tau_s}$. The oscillations get exponentially damped:
\begin{equation} 
{\delta\rho_{xx}\over\rho_0}  =  {\eta^2\over 4} QS [1+2 e^{-\pi Q^2/2S}
\sin 2Q]\ . 
\label{e48}
\end{equation}
\item[iii)] low-field region, $Q\gg Q_2$. The exponential damping
factor changes its form:
\begin{equation} 
{\delta\rho_{xx}\over\rho_0}  =  {\eta^2\over 4} QS [1+2 e^{-\pi/S_s}
\sin 2Q]\ ,
\label{e49}
\end{equation}
where $S_s=\omega_c\tau_s$. Note that Eq.(\ref{e49}) differs from the
low-field behavior for the isotropic scattering case, Eq.(\ref{e24}), 
only in that $\tau$ is replaced by $\tau_s$ in the exponential
damping  factor. 
\end{itemize}
The two regimes (i) and (ii) are jointly described by
\begin{equation}
{\delta\rho_{xx}\over\rho_0}  =  {\eta^2\over 4} Q^2 S
{\pi\over\sinh\pi\mu} J_{i\mu}(Q)J_{-i\mu}(Q)\ ,
\label{e50}
\end{equation}
with $\mu=Q^2/2S$, while both the regimes (ii) and (iii) are given by
\begin{eqnarray} 
{\delta\rho_{xx}\over\rho_0}  &=&  {\eta^2\over 4} QS\left[1+2 
\exp\left\{-{\pi\over S_s} \right.\right.\nonumber \\
&\times & \left.\left.
\left[1-\left(1+{\tau_s\over\tau}Q^2\right)^{-1/2}\right]\right\}
\sin 2Q\right]\ . 
\label{e51}
\end{eqnarray}
Finally all the three regimes can be described by a single formula of
the form (\ref{e50}) with
\begin{equation}
\mu={1\over S_s}
\left[1-\left(1+{\tau_s\over\tau}Q^2\right)^{-1/2}\right]
\label{e52}
\end{equation}

Let us also note that the behavior of $\delta\rho_{xx}$ at $Q\ll 1$
(i.e for magnetic fields stronger than those generating the
oscillations) is the same for all the types of scattering,
\begin{equation}
{\delta\rho_{xx}\over\rho_0} = {\eta^2\over 2} S^2\ ,
\label{e53}
\end{equation}
as is easily found from Eqs. (\ref{e18}), (\ref{e21}),
(\ref{e46}). In reality, in most of the
Weiss oscillation experiments the  magnetoresistivity at such
strong fields is strongly affected by the Shubnikov-de Haas
oscillations and by the quantum Hall effect. However, the giant
quadratic magnetoresistance  (\ref{e53}) was observed in the
experiment \cite{geim}, where the measurements were performed at
somewhat higher temperatures. 

Now we evaluate the boundaries $Q_1$, $Q_2$ of the regions with
different behavior of the oscillation amplitude in a typical
experimental situation. We find from the parameters of the sample
of Ref.\cite{weiss}  ($q=2\pi/382\mbox{nm}$, $n_e=3.16\times
10^{11}\mbox{cm}^{-2}$, $l=12\mu\mbox{m}$) the transport time
$\tau=52\mbox{ps}$. Furthermore, a typical value of the total
relaxation rate for such structures is \cite{coleridge} $\tau_s\simeq
3\mbox{ps}$ (as we will see below, with this value of $\tau_s$ our
results perfectly describe the experimentally observed
 damping of the oscillations in
low fields). This  gives the following crossover values: 
$Q_1\simeq 5.0$, $Q_2\simeq 6.0$. 
For a higher mobility sample studied in Ref.\cite{smet} (the Weiss
oscillation data are presented there in the inset of Fig.~1) these
values are somewhat larger: $Q_1\simeq 8.5$, $Q_2\simeq 11$.
We conclude that although
parametrically $Q_2/Q_1\propto (k_F/q)^{1/3}$ is large at $k_F\gg q$,
for realistic parameters $Q_2$ and $Q_1$ are numerically close to
each other. Therefore, the regime (ii) corresponds in reality to a
rather narrow intermediate domain between
 the regime (i) describing the fast ($\propto
B^3$) drop of $\delta\rho_{xx}$ in high fields, 
and the regime (iii) corresponding to the
$\exp(-\pi/\omega_c\tau_s)$ damping of the oscillatory part of
$\delta\rho_{xx}$ in low fields. 

In Fig.1 we compare our theoretical results with experimental data of
Weiss {\it et al} \cite{weiss}. The sample parameters are \cite{weiss} 
$q=2\pi/382\mbox{nm}$, $n_e=3.16\times
10^{11}\mbox{cm}^{-2}$, $\tau=52\mbox{ps}$, the total relaxation rate
is taken to be $\tau_s^{-1}=(3\mbox{ps})^{-1}$.
As is seen from the figure, 
with the modulation strength  $\eta=0.065$ we get a very good description
of the experimentally observed magnetoresistivity \cite{note2}. 

\begin{figure}
\narrowtext
\centerline{\psfig{figure=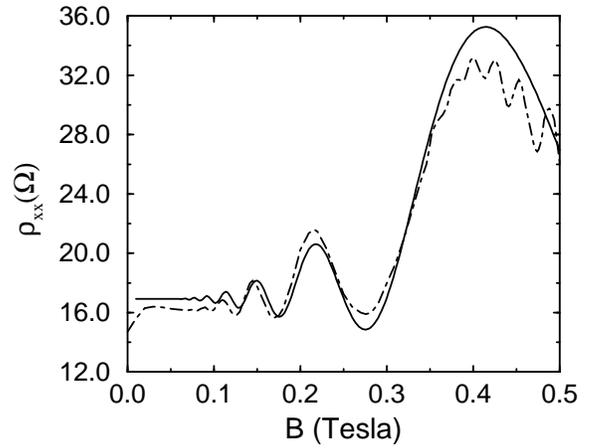,width=9cm}}
\caption{Experimental data of Ref.\protect\cite{weiss} (dash-dotted line)
compared with our theoretical results for the long-range potential
scattering (full line). Parameters are:
$q=2\pi/382\mbox{nm}$, $n_e=3.16\times
10^{11}\mbox{cm}^{-2}$, $\tau=52\mbox{ps}$,
$\tau_s=3\mbox{ps}$, $\eta=0.065$}
\label{fig1}
\end{figure}

To illustrate the
difference between  the long-range potential scattering
and the isotropic scattering, we plotted in Fig.2 the corresponding
results for the modulation-induced resistivity correction
 at the same value of the transport time $\tau$. It is seen that the
results for these two different scattering mechanism
differ drastically not only in low magnetic fields (where the
small-angle scattering leads to the exponential damping of
oscillations), but also in high fields (due to the different power-law 
dependence of $\delta\rho_{xx}$ on $B$). In particular, already
at the first maximum of $\delta\rho_{xx}$, the value of the correction 
is smaller by factor of a $\sim 5$ for the long-range random potential
scattering, as compared to the isotropic scattering case, at the same
value of the grating strength $\eta$. For the second maximum this
ratio is already a factor of $\sim 10$, etc. As a consequence,
one generates a considerable error when trying to evaluate the modulation
strength by fitting the amplitude of one of the first maxima by the
white-noise random potential results. For example, Beenakker
\cite{beenakker} underestimated $\eta$ by factor of 3 \cite{note}.

\begin{figure}
\narrowtext
\centerline{\psfig{figure=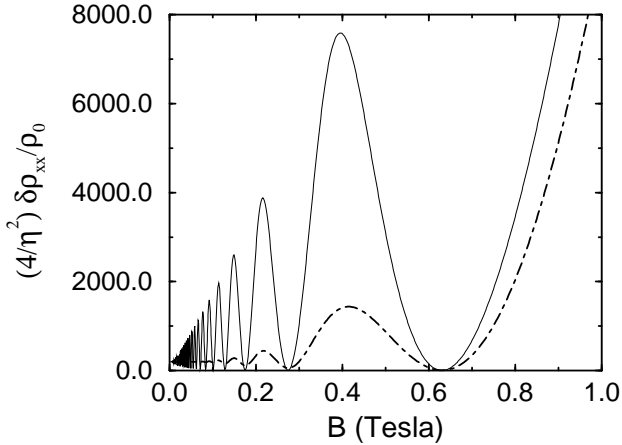,width=9cm}}
\caption{Grating-induced correction to the 2DEG reistivity,
$\delta\rho_{xx}/\rho_0$, in units of $\eta^2/4$ for the isotropic
potential scattering ($\tau=52\mbox{ps}$, full line) and the
long-range random potential scattering ($\tau=52\mbox{ps}$,
$\tau_s=3\mbox{ps}$, dash-dotted line). The sample parameters
($q=2\pi/382\mbox{nm}$, $n_e=3.16\times
10^{11}\mbox{cm}^{-2}$)   are the
same as in Ref.\protect\cite{weiss} and in Fig.1.} 
\label{fig2}
\end{figure}

\section{Conclusions}
\label{s7}

In this paper, we have presented a theory of the Weiss oscillations of
the magnetoresistivity in a 2DEG, for the case
when the impurity scattering is
predominantly of the small-angle nature. 
While this is precisely the experimental situation in currently
fabricated semiconductor heterostructures, 
no quantitative theoretical description of this case has been
available so far.
Previous theories \cite{gerhardts89,winkler,beenakker} assumed
the scattering to be isotropic. We have shown that the small-angle
scattering strongly affects the shape and the damping of the Weiss
oscillations. This was demonstrated both for the scattering by a
white-noise random magnetic field (linearly increasing
eigenvalues $\lambda_l=\tau^{-1}(2|l|-1)$ of the collision integral)
and for  long-range random potential scattering ($\lambda_l\simeq
\tau^{-1}l^2/[(\tau_s/\tau)l^2+1]$).
 In the latter case, which was of primary interest for us, the shape
of the magnetoresistivity  is determined by both the
momentum and the total relaxation rates ($\tau^{-1}$ and
$\tau_s^{-1}$, respectively), and has the following features. In
relatively strong magnetic fields $B$, the amplitude of the modulation
induced correction $\delta\rho_{xx}$ falls of as $B^3$ with decreasing
$B$, i.e. much more rapidly, than in a white-noise random potential,
where  $\delta\rho_{xx}\propto B$. This rapid fall-off is determined by
the quadratic increase of the eigenvalues $\lambda_l\propto l^2$ for
not too large $l$. In low magnetic fields, the oscillatory part of 
the magnetoresistivity is found to be exponentially damped, much more
efficiently than in the case of isotropic scattering with the same
$\tau$. The low-field damping factor is of the form
$e^{-\pi/\omega_c\tau_s}$, familiar from the theory
 of the Shubnikov-de Haas oscillations.  Let us stress, however, 
that while the Shubnikov-de Haas oscillations are of purely quantum
nature, the Weiss oscillations represent a classical
effect. Nevertheless, due to the interplay of the finite modulation wave
vector $q$ and the cyclotron radius $R_c$, the resistivity correction
$\delta\rho_{xx}$ is determined by the whole spectrum of eigenvalues
$\lambda_l$ of the collision integral (and, in particular, by
$\lambda_\infty=1/\tau_s$), rather than by $\lambda_1=1/\tau$ only 
(governing conventional transport in the absence of modulation). 
Finally, our results describe nicely the shape of the experimentally
observed magnetoresistance and the dependence of the oscillation
amplitude on the magnetic field. 

We are grateful to D.~E.~Khmelnitskii for attracting our attention to
the problem of Weiss oscillations and to
Y.~Levinson for useful discussions at various 
stages of this work. 
This work was supported by the SFB195 der Deutschen
Forschungsgemeinschaft.

\end{multicols}

\begin{references}
\bibitem{note} Note that the r.m.s. amplitude $\epsilon$
of the density modulation used by Beenakker \cite{beenakker}
is related to our parameter $\eta$ as $\epsilon=\eta/\sqrt{2}$.
\end{references}
\end {document}